# Original Ti:Sa 10 kHz Front-end design delivering 17 fs, 170 mrad CEP stabilized pulses up to 5 W

A. Golinelli[1,2,*], X. Chen[1], E. Gontier[1], B. Bussière[1], O. Tcherbakoff[2], M. Natile[1,2], P. d'Oliveira[2], P.-M. Paul[1,3] and J.-F. Hergott[2]

[1] Amplitude Technologies, 2 - 4 rue du Bois Chaland CE 2926, 91029 Evry, France
[2] LIDYL, CEA, CNRS, Université Paris-Saclay, CEA-SACLAY, 91191 Gif-sur-Yvette, France
[3] Continuum Inc., Amplitude Laser Group, 140 Baytech Drive, CA 95134, San Jose, USA
*Corresponding author: agolinelli@amplitude-technologies.com

**We present a compact 10 kHz Ti:Sa front-end relying on an original double-crystal regenerative amplifier design. This new configuration optimizes the thermal heat load management, allowing producing a 110 nm large spectrum and maintaining a good beam profile quality. The front-end delivers up to 5 W after compression, 17 fs pulses with a 170 mrad shot-to-shot residual CEP noise. © 2015 Optical Society of America**

The latest efforts devoted to a deeper understanding of attosecond temporal dynamics have led to a rapid development of ultrafast laser technology. High energy, few cycle laser pulses represent the main tool for accessing the strong-field interactions that rule these dynamics. When an ultrashort, high-intensity (above $10^{14}$ W/cm$^2$) driving pulse interacts with a medium, non-linear effects take place, producing a train of XUV attosecond pulses through high harmonic generation [1, 2]. For many applications [3], like pump-probe experiment [4], nano-plasmonic [5], molecular vibrational wavepacket mapping [6], atomic correlation investigation [7] and attosecond lighthouse [8], a single isolated attosecond pulse rather than a train of attosecond pulses is needed. Confining the emission of XUV radiation to one half of the optical cycle near the peak of the driving pulse is the straightforward mean to obtain isolated attosecond pulses. This confinement has been obtained by post-compression [9, 10] or polarization gating [11, 12] of high energy, sub-20 fs pulses. As the pulse duration approaches the single cycle regime, the control of the relative Carrier-Envelope Phase – CEP – becomes crucial for a faithful replicability of the isolated attosecond pulse generation process. The CEP stability of the driving pulse, as well as ultra-short pulse duration, is a key parameter for successful generation of isolated attosecond pulses. The efficiency of the high harmonic generation process sets additional requirements for the driving source: high repetition rate and high energy pulses. These features ensure a meaningful brightness of XUV signal while reducing the acquisition time.

A grating-based CPA system has been demonstrated to be, since now, the most suitable choice for achieving high-energy pulses, keeping the B-integral value low, for Ti:Sa based amplifier [13]. On the other hand, reflection gratings are notably more sensitive to mechanical vibration: the resulting beam pointing variation in stretcher and compressor causes dispersion fluctuations and thus CEP shift. In this context the crucial parameters to CEP noise are the distance and mutual angular alignment of the gratings in the stretcher and compressor. It has been demonstrated [14, 15, 16] that small vibrations and movements of the gratings induce a relevant CEP shift.

Besides mechanical vibrations, two other factors inside the amplifier influence the CEP stabilization [17]. One is the path length change of laser beam linked to air turbulence or mechanical vibration as well as material temperature fluctuation. The influence of typical beam depointing to shot-to-shot CEP noise is on sub-mrad level; the effect on CEP from media temperature variation can be several orders greater than beam depointing, but it occurs on a few minutes time scale. Generally, beam depointing and temperature fluctuation introduce a negligible CEP noise for good environmental conditions. The second factor is laser intensity fluctuations of the amplified pulses due to a refractive index gradient in the amplification crystal. Both pump power fluctuations and overlap mismatching between pump and seed beam, can cause abrupt refractive index variation. At the same time, pump energy stability and fast beam depointing introduced by mechanical vibrations in amplifiers also play an important role on CEP stabilization. Typical spectral distributions of CEP noise are up to 1 kHz for mechanical vibrations in grating-based modules and above 10 kHz for intensity fluctuations [18]. While earlier spectral interferometers limited the acquisition speed to milliseconds [19], latest digital detectors can reach kHz acquisition rate [18]. Besides the detection speed, the actuation bandwidth also plays an important role for an efficient CEP stabilization. Feedback loops to actuators acting on mechanical part displacement, e.g. grating distances [14] or glass wedges insertion [20], limit the actuation loop to few Hertz. Fast detection system combined with fast actuator is therefore necessary for a better stabilization of the CEP.

We present here an original, full water-cooled front-end system which is capable of delivering CEP stabilized, 17 fs pulses at 10 kHz repetition rate. The output energy of the compressed pulses is around 500 micro joule. Our technology is based on the Chirped Pulse Amplification technique [21], relying on a Ti:Sa regenerative amplifier in a double-crystal design. The schematic experimental setup of the laser is shown in Figure 1.

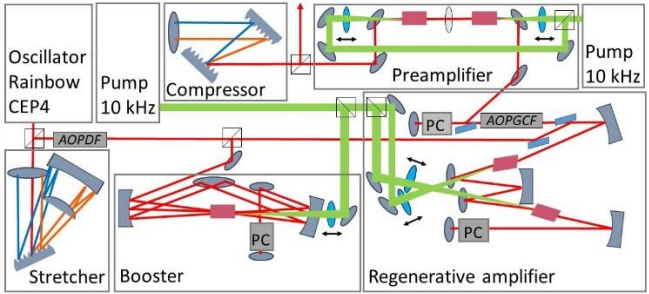

Fig. 1. Scheme of the experimental setup in CPA configuration. The oscillator is CEP stabilized (Rainbow CEP4, Femtolaser). The Öffner-triplet stretcher is followed by three amplification stages and by a compressor. The amplifiers are pumped by two 10 kHz Continuum MESA lasers, with 150 ns pulse duration and a $M^2$ factor of 25. PC – Pockels cell.

Double crystal regenerative amplifiers have already been demonstrated for thermal load repartition at 1 kHz repetition rate in Yb:KYW [22] and Yb:CALGO crystals [23]. A double-slab configuration amplifier has also been designed for gain narrowing compensation in a multi-kHz (from 20 kHz to 100 kHz) regenerative amplifier [24]. The idea behind this configuration is to combine together the different emission spectra of two Yb:KYW crystals, aligned according to two different optical axes. Applying the double crystal design, for the first time to our knowledge, to a Ti:Sa crystal-based cavity, allows to manage thermal lensing.

The system has been developed by Amplitude Technologies in collaboration with CEA Saclay within the joint laboratory Impulse. The front-end is seeded by a CEP stable oscillator (Rainbow CEP4, Femtolaser) producing pulses at 75 MHz repetition rate with a residual theoretical CEP noise below 60 mrad [25]. A grating-based (1200 grooves/mm) stretcher in an Öffner-triplet configuration temporally extends the pulses with a stretching factor of 5 ps/nm with 150 nm spectral bandwidth throughput. This allows high amplification while keeping a low B-integral value. To ensure an appropriate compression of the output pulse, the Öffner stretcher is coupled to a grating-based compressor.

A Dazzler module (AOPDF) is placed at the output of the stretcher [26]. When feedback by phase and CEP measurement systems, the Dazzler can provide large range dispersion compensation, in addition to the compressor, and CEP slow drift correction at the same time.

A 6 pass amplifier boosts the pulses energy before further amplification in the regenerative amplifier. A KD*P Pockels Cell between the first three and the last three passes decreases the repetition rate to 10 kHz. The ring configuration of this multipass amplifier, or booster, ensures small angles between each pass and the pump inside the Ti:Sa crystal; this good superposition between the pump (15 W) and the IR beam allows an overall gain of 200, leading to an IR pulse energy of 50 nJ. Increasing the seed energy helps to obtain a better contrast between the picosecond and the nanosecond regime after the regenerative cavity.

The pulses are then amplified in a double Brewster-cut Ti:Sa regenerative cavity. The crystals length is 20 mm; the doping concentration ensures 90% absorption of the incident pump power. This new W-like designed cavity generates high energy, broadband pulses whose duration can be shortened below 20 fs. Ultrashort time duration requires wide spectrum. Unfortunately, this is limited by gain narrowing occurring in the Ti:Sa crystal amplification[27]. This effect can be counteracted by means of an intracavity programmable filter (AOPGCF or Mazzler) [28] which shapes the spectral losses of the cavity to ensure a homogeneous amplification over a large spectrum. Spectral losses are created where the spectral gain is higher, leading to an amplification enhancement at the edges of the spectrum. The losses introduced by the Mazzler are compensated by increasing the pump power. However, pumping with high average power at 10 kHz repetition rate causes thermal heat load accumulation in the crystal [29] giving rise to two main phenomena: refractive index gradient and thermo-mechanic deformation of the crystal. These two phenomena contribute together to increase the thermal lens inside the crystal, whose focal length is described by the following formula [30, 31]:

$$f_{th} = \frac{2\pi\, r_1^2\, K_1}{P_{th}\left[\left(\frac{dn}{dT}\right)_{T=T_0} + \frac{2\, r_2\, \alpha_{th}\, (n_0 - 1)}{L}\right]}$$

where $r_1$ is the radius of the pumped zone on the crystal, $r_2$ is the radius of the crystal, $K_1$ is the thermal conductivity, $dn/dT$ is the refractive index gradient in function of the temperature, $\alpha_{th}$ is the coefficient of thermal expansion, $L$ is the length of the crystal, $P_{th}$ is the fraction of pump power converted in thermal power.

Even though proper dimensioning of the crystal and cooling systems help reducing the heat load, the thermal effect is critically influenced by the pump power density.

As the pump power increases, the focal length of the thermal lens becomes shorter, leading to instability of the cavity, aberration and thus poor beam profile. At optimal pump power the focal length of the thermal lens can be as short as ten centimeters. Figure 2.a. shows the beam profile evolution for increasing pump power in a single crystal cavity. In this standard configuration the requirement of a clean beam profile limits the pump power to 14 W, thus limiting the output power to 1.8 W. The new developed double-crystal cavity is able to overcome the power limit caused by thermal effects. Sharing the pump power and thermal heat load in two crystals, so that each crystal is pumped by approximately 14 W, allows the double crystal cavity to maintain a good beam profile quality for pulses whose energy is up to three times higher than with a standard single crystal cavity. Each crystal is pumped with 2 J/cm$^2$ energy density and the pump beam diameter matches the one of the seed. Injection and extraction of the amplified beam is performed by two separated KD*P Pockels cells. Astigmatism resulting from the Brewster cut crystals and thermal lens spatial inhomogeneity is compensated by fine tuning of the angle of the spherical mirrors. Figure 2.b. shows the beam profile evolution as function of total pump power inside the cavity. We can see that the double-crystal cavity remains stable over a large range of pump power, which corresponds to a large range of thermal lensing.

Pumped with a total power of 27 W, this double Ti:Sa crystal cavity could generate pulses up to 5.1 W (0.3 % RMS stability) in moderate spectral bandwidth, i.e. 50 nm at $1/e^2$. When AOPGCF is implemented and after optimization process an amplified spectrum up to 110 nm at $1/e^2$ is obtained with an output power of 2.6 W (0.3 % RMS stability) for 28.5 W of pump. As comparison, a single crystal cavity pumped at optimal condition for excellent beam profile with 80 nm spectral width would deliver less than 1 W output power (we were unable to reach 110nm spectral width in this configuration). Thus the double-crystal cavity is about three times more efficient. Reaching higher output power in a single crystal cavity is possible with increasing the pump power at the cost of a poor beam profile. This effect in turn causes the cavity to be unstable and the Mazzler to be less efficient with higher losses for spectral broadening.

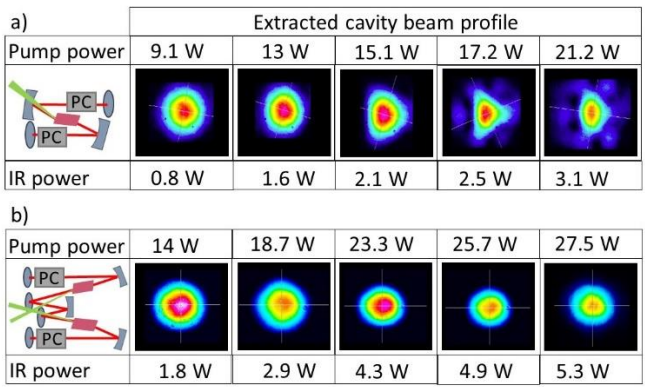

Fig. 2. Evolution of the beam profile and IR output power as a function of the total pump power seeded in (a) single crystal and (b) double crystal regenerative amplifier.

Further amplification up to 7 W (0.35 % RMS stability) is achieved in a two-crystal single pass amplifier (preamplifier) pumped with 25 W. Since the single pass amplifier is also affected by gain narrowing and red-shifting of the amplified pulse, the spectral shaping via Mazzler is performed using the output spectrum of the preamplifier. The gain narrowing pre-compensation ensures that the spectrum at the output of the preamplifier has the desirable width and shape for a pulse duration as short as possible. For the pumping scheme of all amplifiers dichroic mirrors have been used to ensure high transmission at the pump wavelength and high reflectivity at the IR wavelength. In the meantime pump and IR beams paths are adjusted in order to ensure a good overlap inside the crystals.

The last module of the front-end is the reflection grating-based (1480 grooves/mm) compressor. To prevent our system from being critically sensitive to mechanical vibrations, considering also the extended dimension, new CEP stable grating mounts have been designed for the stretcher and for the compressor. The new mounts are without rotation and translation stages, in order to ensure a solid and stable support for the optics.

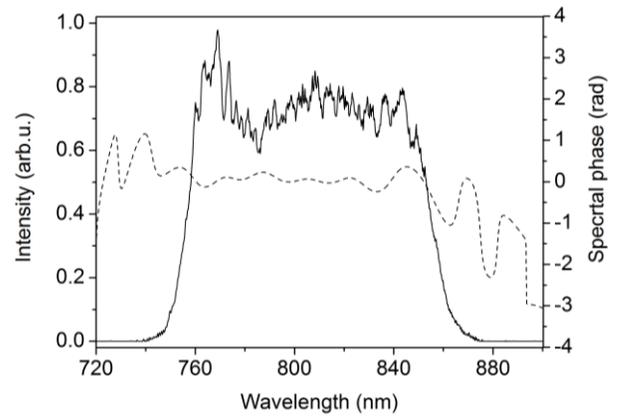

Fig. 3. Measured 110 nm broad spectrum at $1/e^2$ (black line) and spectral phase (dashed line).

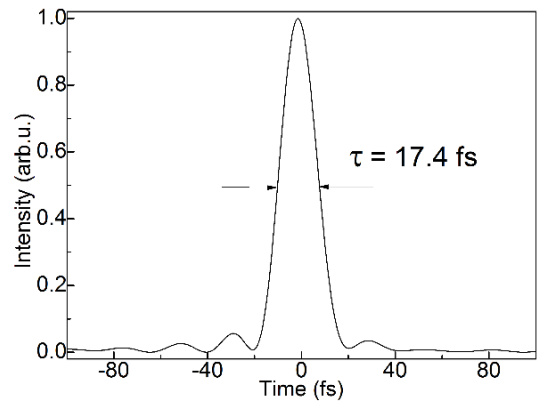

Fig. 4. Measured temporal profile after optimized pulse compression corresponding to 17.4 fs (Fourier transform limited pulse duration is 17 fs).

Figure 3 shows the spectrum of the pulses: 110 nm spectrum corresponds to a Fourier transform limited pulse duration of 17 fs. Thanks to the high order dispersion compensation performed by the Wizzler-Dazzler feedback loop, the pulse duration is as short as 17.4 fs, as shown in figure 4.

The residual CEP noise of the amplified pulses has also been measured. The CEP stabilization method proposed here is based on a complete analog acquisition system coupled to a fast actuator [32]. More specifically, we perform simultaneously an analog and a digital acquisition. The interference fringes produced by a home-made f-to-2f interferometer are sent, by means of a beam splitter, to two arms. One arm is a digital fast spectrometer for long time measurements. The spectrometer resolves spectrally the interference fringes and their displacement due to CEP noise. The digital acquisition is limited to 1 kHz acquisition rate. The other arm is the analog acquisition running at the full laser repetition rate and coupled with the correction loop. Two photomultipliers detect the spatial displacement of the interference fringes correlated to shot-to-shot CEP noise and commute it into an error signal. The latter is sent to a PID controller generating a voltage output signal which drives the actuator, i.e. the Dazzler. Since no mechanical part displacement is needed for the feedback loop, the actuator is characterized by a fast response.

Figure 5.b. shows the phase noise Power Spectral Density of a 20 seconds CEP measurement performed with the fast oscilloscope, for analog amplifier feedback loop ON and OFF. The residual shot-to-shot CEP fluctuations is as low as 170 mrad (standard deviation). The combination of the analogic detection and the fast stabilization loop allows a significant correction of the CEP drift up to 200 Hz. Figure 5.a. shows long time measurements performed with the spectrometer over 3 hours: CEP stabilization is reached with a residual shot-to-shot noise as low as 210 mrad. This long time CEP stabilization detection is performed at 1 kHz with the digital acquisition leading to a sampled shot-to-shot CEP measurement (1 shot over 10). Integrated over 10 shots, the CEP residual noise is lowered down to 110 mrad.

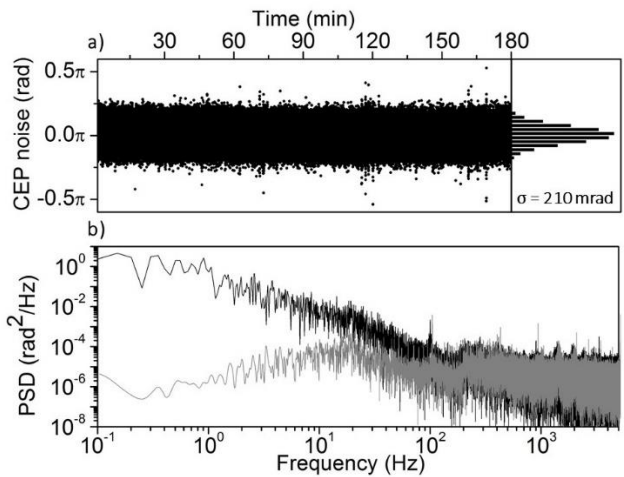

Fig. 5. a) CEP measurement performed with the spectrometer over 3 hours; the residual CEP noise is 210 mrad. b) Power Spectral Density of the CEP variation when oscillator is CEP stabilized and (black) analog feedback loop is OFF or (gray) analog feedback loop is ON. The CEP residual noise with analog feedback loop ON is as low as 170 mrad.

In conclusion, we demonstrate a novel regenerative amplifier configuration capable of managing thermal effects occurring at high repetition rate. The front-end delivers pulses with broad spectrum corresponding to 17 fs measured pulse duration and 5 W output power after compression. CEP stabilization of the 10 kHz front-end has also been performed, by means of the combination of a home-made analog fast detector and a fast actuator (DAZZLER). Residual shot-to-shot CEP noise has been measured as low as 210 mrad over several hours, confirming the reliability of the system over a long running time. To our knowledge, those are the best results reported for a Ti:Sa laser system with grating-based stretcher and compressor. Allowing good beam profile quality and high output power at the first amplification stage, in comparison with a classical design, the innovative double-crystal regenerative cavity is one of the most promising configurations for future high intensity CEP stable lasers.

**Funding.** European Union H2020-MSCA-ITN-MEDEA-641789; Agence Nationale de la Recherche ANR11-EQPX0005 ATTOLAB.